\documentclass[aps,prd,letterpaper,showkeys,showpacs,onecolumn,nofootinbib,preprintnumbers,longbibliography,unsortedaddress]{revtex4-1}

\usepackage[top=2.8cm, bottom=2.8cm, left=2cm, right=2cm]{geometry}
\usepackage[utf8]{inputenc}

\usepackage{amsmath,amssymb,lmodern,amsfonts} 
\usepackage{bm,latexsym,amsmath,amssymb,amsfonts,mathrsfs,ulem}
\usepackage[dvipdfmx]{graphicx}
\usepackage{natbib}
\usepackage{hyperref} 
\usepackage{color}
\usepackage{tabulary} 
\usepackage{empheq}
\usepackage[x11names]{xcolor}

\newcommand{\be}{\begin{equation}}
\newcommand{\ee}{\end{equation}}
\newcommand{\bea}{\begin{eqnarray}}
\newcommand{\eea}{\end{eqnarray}}

\def\ba{\begin{array}}
\def\ea{\end{array}}
\def\bfig{\begin{figure}}
\def\efig{\end{figure}}

\def\beq{\begin{equation}}
\def\eeq{\end{equation}}
\def\bi{\begin{itemize}}
\def\ei{\end{itemize}}
\def\ba{\begin{array}}
\def\ea{\end{array}}
\def\bfig{\begin{figure}}
\def\efig{\end{figure}}

\newcommand{\eqn}[1]{(\ref{#1})}
\newcommand{\eq}[1]{Eq.~(\ref{#1})}

\begin{document}
\newcommand{\agc}[1]{{\color{red} #1}}

\preprint{PI/UAN-2022-714FT}

\title{Decoupling-limit consistency of the  generalized SU(2) Proca theory}

\author{Alexander Gallego Cadavid}
\email{alexander.gallego@uv.cl}
\affiliation{Instituto de F\'{\i}sica y Astronom\'{\i}a,  Universidad  de Valpara\'{\i}so, \\ Avenida Gran Breta\~na 1111,  Valpara\'{\i}so  2360102,  Chile}

\author{Carlos M. Nieto}
\email{caniegue@correo.uis.edu.co}
\affiliation{Escuela  de  F\'isica,  Universidad  Industrial  de  Santander, \\ Ciudad  Universitaria,  Bucaramanga  680002,  Colombia}

\author{Yeinzon Rodr\'iguez}
\email{yeinzon.rodriguez@uan.edu.co}
\affiliation{Centro de Investigaciones en Ciencias B\'asicas y Aplicadas, Universidad Antonio Nari\~no, \\ Cra 3 Este \# 47A-15, Bogot\'a D.C. 110231, Colombia}
\affiliation{Escuela  de  F\'isica,  Universidad  Industrial  de  Santander, \\ Ciudad  Universitaria,  Bucaramanga  680002,  Colombia}

\begin{abstract}
We study the consistency of the decoupling limit of the generalized SU(2) Proca theory (GSU2P). Namely, we study the healthiness of those terms whose analysis in the scalar limit was not originally established in the reconstruction of the full theory (see the work by Gallego Cadavid {\it et. al.} [Phys. Rev. D 102, 104066 (2020)]).  Those terms are the parity-violating $\tilde{\mathcal{L}}_{4,2}^1$ and the parity-conserving beyond SU(2) Proca terms $\mathcal{L}_{4,2}^3$ and $\mathcal{L}_{4,2}^4$. Using the 3+1 Arnowitt-Deser-Misner formalism, we write down the kinetic Lagrangian of these terms in the decoupling limit and show that their corresponding kinetic matrices are degenerate. This degeneracy is a necessary condition for the propagation of the right number of degrees of freedom, as required by the primary constraint-enforcing relation.  Interestingly, the $\tilde{\mathcal{L}}_{4,2}^1$ term, which is purely non-Abelian, does not contribute to the kinetic Lagrangian of the theory, so its contribution is trivially degenerate. Similarly, but not trivially in these cases, the contributions of the $\mathcal{L}_{4,2}^3$ and $\mathcal{L}_{4,2}^4$ terms to the kinetic Lagrangian turn out to be degenerate as well. The results presented in this paper represent progress
in the construction of the fully healthy GSU2P.

\end{abstract}



\maketitle

\section{Introduction}
Lovelock's theorem \cite{Lovelock:1971yv,Lovelock:1972vz} gave us a very strong message:  general relativity (GR) is the unique gravity theory in four dimensions, written in terms of the metric and up to its second-order derivatives, that is free of instabilities because its field equations are of second order.  However, we always have in mind the possibility of modifying GR because it is an effective theory \cite{Donoghue:1994dn,Burgess:2003jk} (unless it can be shown that it is non-perturbatively renormalizable \cite{Reuter:2001ag,Litim:2003vp,Falls:2014tra,Eichhorn:2018yfc}).  Therefore, to construct modifications of gravity, we need to find a way to avoid Lovelock's theorem. And the way to avoid a theorem is by relaxing one or more of the hypotheses which it is laid on.  

The literature on gravity theories is full of papers that break, in some way or the other, the body of hypotheses of Lovelock's theorem.  One of the taken avenues consists of adding more gravitational degrees of freedom in the form of scalar, fermion, vector, or rank-2 tensor fields.  Perhaps, the most significant work in such a direction is that of Horndeski's, who built in the 70's the most general scalar-tensor field gravity theory whose field equations are second order \cite{Horndeski:1974wa}.  Since then, the line of thought of this work has been followed in order to replace the scalar field either by: a vector field enjoying a U(1) gauge symmetry \cite{Horndeski:1976gi}, multiple scalar fields \cite{Padilla:2012dx,Sivanesan:2013tba,Padilla:2010ir,Allys:2016hfl} (the multi-Galileons theory), a vector field free of gauge symmetries \cite{Tasinato:2014eka,Heisenberg:2014rta,Allys:2015sht,Jimenez:2016isa,Allys:2016jaq} (the generalized Proca theory (GP)), or a vector field enjoying a global SU(2) symmetry \cite{GallegoCadavid:2020dho,Allys:2016kbq,Jimenez:2016upj} (the generalized SU(2) Proca theory (GSU2P)), among other possibilities\footnote{The Horndeski theory was forgotten for many years until it was rediscovered around 2010 \cite{Nicolis:2008in,Deffayet:2009mn,Deffayet:2009wt,Deffayet:2011gz,Deffayet:2013lga}.  The equivalence between the new ``Galileon'' theory and that of Horndeski's was established in Ref. \cite{Kobayashi:2011nu}.  For nice reviews see Refs. \cite{Kobayashi:2019hrl,Heisenberg:2018vsk,Rodriguez:2017ckc}.}.

All the previously mentioned theories follow the idea of producing second-order field equations, both in the full theory and in its decoupling limit, to avoid the Ostrogradski's instability \cite{Ostrogradsky:1850fid}.  This instability refers to the situation when there is no ground state in a physical system \cite{Woodard:2006nt,Woodard:2015zca}. However, the second-order nature of the field equations is just a sufficient, but not a necessary, condition to avoid the Ostrogradski's instability.  Indeed, it was later recognized that such an instability, which is intimately connected to the propagation of unphysical degrees of freedom, could be appropriately removed by imposing a high enough number of constraints in the Hamiltonian that are conserved in time (following Dirac's Hamiltonian constraints algorithm \cite{Dirac:1964,Blagojevic:2002du}).  The key, therefore, was the kinetic matrix whose degeneracy becomes the primary constraint-enforcing relation.  Thus, several theories of higher-order nature were born:  the beyond Horndeski theory \cite{Zumalacarregui:2013pma,Gleyzes:2014dya}, the extended scalar-tensor theory (or degenerate higher-order scalar-tensor theory (DHOST) \cite{Langlois:2015cwa,Achour:2016rkg,BenAchour:2016fzp,Crisostomi:2016czh}), the beyond Proca theory \cite{Heisenberg:2016eld,GallegoCadavid:2019zke}, the beyond SU(2) Proca theory \cite{GallegoCadavid:2020dho}, the extended vector-tensor theory \cite{Kimura:2016rzw}, and the extended SU(2) Proca theory \cite{GallegoCadavid:2021ljh}.

It is very important to highlight that the full versions of the GP and the GSU2P are degenerate by construction to suppress the propagation of a fourth degree of freedom in the vector fields\footnote{The temporal component of each vector field, to be precise.} (incompatible with the Poincaré group representations).  Additionally, the decoupling limit of the GP as well as the same limit of most of the Lagrangian pieces that compose the GSU2P follow the idea of producing second-order field equations.

The GSU2P was built in Ref. \cite{Allys:2016kbq}, implementing the degeneracy constraint for the propagation of the right number of degrees of freedom.  This constraint, however, is not the end of the story:  the constraints algebra can be multi-dimensional which means that a gravity theory must be constructed by having in mind that the other $n-1$ constraints, $n$ being the dimension of the constraints algebra, must be satisfied as well.  Unfortunately, neither the GP nor the GSU2P was constructed by having in mind this consideration.  Indeed, Refs. \cite{ErrastiDiez:2019trb,ErrastiDiez:2019ttn} showed that, for Maxwell-Proca theories, the constraints algebra is two-dimensional.  The secondary constraint turned out to be trivially satisfied by the already constructed GP but not by the GSU2P.  A delicate analysis of the GSU2P constructed in Ref. \cite{Allys:2016kbq} revealed that many terms were discarded as redundant via total derivatives in the Lagrangian that did not satisfy the secondary constraint.  Of course, this ruined the entire construction, forcing the rebuilding of the theory from scratch.  This was properly done in Ref. \cite{GallegoCadavid:2020dho} with the additional advantage of getting the beyond SU(2) Proca terms following the technique devised in Ref. \cite{GallegoCadavid:2019zke}. The reconstructed GSU2P in Ref. \cite{GallegoCadavid:2020dho} is composed of fourteen Lagrangian pieces, eight of them being parity conserving, other five being parity violating, and the other one being a mixture of both parity-conserving and parity-violating terms.  The terms belonging to the beyond SU(2) Proca theory are included in this list, they being five of the fourteen, two parity conserving and three parity violating.  The decoupling limits were analyzed for the fourteen pieces, and it was found that eleven of them either vanish or lead to second-order field equations for the scalar fields representing the longitudinal modes of the vector fields. The remaining three pieces:  the purely non-Abelian\footnote{Here, ``purely non-Abelian" means that it vanishes when stripped out of the internal group indices.} and parity-violating $\tilde{\mathcal{L}}_{4,2}^1$ and the two parity-conserving beyond SU(2) Proca terms $\mathcal{L}_{4,2}^3$ and $\mathcal{L}_{4,2}^4$ lead to higher-order field equations that put their healthiness into question.  The purpose of this paper is to check whether the decoupling limits of $\tilde{\mathcal{L}}_{4,2}^1$, $\mathcal{L}_{4,2}^3$, and $\mathcal{L}_{4,2}^4$ satisfy the degeneracy condition and, therefore, implement the primary constraint.  As we will see, the answer is positive:  they do pass the test.

The layout of the paper is as follows.  In Section \ref{bgsu2pt}, we present the decoupling limits of the three Lagrangian pieces under consideration. In Section \ref{decomp}, we briefly review the 3+1 Arnowitt-Deser-Misner (ADM) decomposition (see Ref. \cite{3+1book}) in the decoupling limit for a general Lagrangian containing an SU(2) vector field $B_{\mu}^a$. Then, we perform a basis transformation in order to simplify the kinetic Lagrangian and block-diagonalize the corresponding kinetic matrix. In Section \ref{det}, we apply the former procedure to each of the Lagragian pieces of interest. Here, we show that the respective kinetic matrices of the beyond SU(2) Proca terms are degenerate and that the respective kinetic Lagrangian of $\tilde{\mathcal{L}}_{4,2}^1$ is identically zero. Finally, the conclusions are presented in Section \ref{conclusions}.

Throughout the text, we use Greek indices as space-time indices which run from 0 to 3, while the first letters of the Latin alphabet label the internal SU(2) group indices. On the other hand, the letters $i$, $j$ and $k$ are used to label the vectors in the two basis introduced in the text;  all these Latin indices run from 1 to 3.  The sign convention is the (+++) according to Misner, Thorne, and Wheeler \cite{Misner:1974qy}. We set $B_{\mu}^a B^{\mu}_a \equiv  B^a \cdot B_a \equiv B^2$.

\section{The GSU2P} \label{bgsu2pt}
As discussed in the introduction, the GSU2P is composed of fourteen Lagrangian pieces that propagate the right number of degrees of freedom (three among four for a vector field) and produce second-order field equations for the vector fields. It is described by the following action\footnote{The conventions have been changed a bit with respect to those in Ref. \cite{GallegoCadavid:2020dho} to increase clarity:  there, the symbol $A$ has been employed to denote both the vector field and the antisymmetric part of twice its covariant derivative;  here, the symbol $A$ is reserved for the mentioned antisymmetric part while the symbol $B$ now denotes the vector field.} (to see complete details about its construction, see Ref. \cite{GallegoCadavid:2020dho}): 
\begin{equation}
S = \int d^4 x \ \sqrt{-g} \ \left(\mathcal{L}_{E-H} + \mathcal{L}_2 + \alpha_{4,0} \mathcal{L}_{4,0} + \mathcal{L}_{4,2} + \frac{\alpha_{5,0}}{m_P^2} \tilde{\mathcal{L}}_{5,0} \right) \,, \label{GSU2Paction}
\end{equation}
where
\begin{eqnarray}
\mathcal{L}_{E-H} &\equiv& \frac{m_P^2}{2} R \,, \label{EH} \\
\mathcal{L}_2 &\equiv& \mathcal{L}_2(A_{\mu \nu}^a,B_\mu^a) \,, \\
\mathcal{L}_{4,0} &\equiv& G_{\mu \nu} B^{\mu a} B^\nu_a \,, \\
\mathcal{L}_{4,2} &\equiv& \sum_{i = 1}^6 \frac{\alpha_i}{m_P^2} \mathcal{L}_{4,2}^i + \sum_{i = 1}^4 \frac{\tilde{\alpha}_i}{m_P^2} \tilde{\mathcal{L}}_{4,2}^i \,, \label{L42} \\
\tilde{\mathcal{L}}_{5,0} &\equiv& B^{\nu a} R^\sigma_{\ \ \nu \rho \mu} B_\sigma^b \tilde{A}^{\mu \rho c} \epsilon_{abc} \,, \label{L50}
\end{eqnarray}
and
\begin{eqnarray}
\mathcal{L}_{4,2}^1 &\equiv& (B_b \cdot B^b) [S^{\mu a}_\mu S^\nu_{\nu a} - S^{\mu a}_\nu S^\nu_{\mu a}] + 2 (B_a \cdot B_b) [S^{\mu a}_\mu S^{\nu b}_\nu - S^{\mu a}_\nu S^{\nu b}_\mu] \,, \nonumber \\
\mathcal{L}_{4,2}^2 &\equiv& A_{\mu \nu}^a S^{\mu b}_\sigma B^\nu_a B^\sigma_b - A_{\mu \nu}^a S^{\mu b}_\sigma B^\nu_b B^\sigma_a + A_{\mu \nu}^a S^{\rho b}_\rho B^\mu_a B^\nu_b \,, \nonumber \\
\mathcal{L}_{4,2}^3 &\equiv& B^{\mu a} R^\alpha_{\ \ \sigma \rho \mu} B_{\alpha a} B^{\rho b}  B^\sigma_b + \frac{3}{4} (B_b \cdot B^b) (B^a \cdot B_a) R \,, \nonumber \\
\mathcal{L}_{4,2}^4 &\equiv& [(B_b \cdot B^b) (B^a \cdot B_a) + 2 (B_a \cdot B_b) (B^a \cdot B^b)] R \,, \nonumber \\
\mathcal{L}_{4,2}^5 &\equiv& G_{\mu \nu} B^{\mu a} B^\nu_a (B^b \cdot B_b) \,, \nonumber \\
\mathcal{L}_{4,2}^6 &\equiv& G_{\mu \nu} B^{\mu a} B^{\nu b} (B_a \cdot B_b) \,, \label{L42LPa}
\end{eqnarray}
\begin{eqnarray}
\tilde{\mathcal{L}}_{4,2}^1 &\equiv& -2 A_{\mu \nu}^a S^{\mu b}_\sigma B_{\alpha a} B_{\beta b} \epsilon^{\nu \sigma \alpha \beta} + S^a_{\mu \nu} S^{\nu b}_\sigma B_{\alpha a} B_{\beta b} \epsilon^{\mu \sigma \alpha \beta} \,, \nonumber \\
\tilde{\mathcal{L}}_{4,2}^2 &\equiv& A_{\mu \nu}^a S^{\mu b}_\sigma B_{\alpha a} B_{\beta b} \epsilon^{\nu \sigma \alpha \beta} - \tilde{A}^{\alpha \beta}_a S^b_{\rho \alpha} B^{\rho a} B_{\beta b} + \tilde{A}^{\alpha \beta}_a S^\rho_{\rho b} B_\alpha^a B_\beta^b \,, \nonumber \\
\tilde{\mathcal{L}}_{4,2}^3 &\equiv& B_\beta^b R^\alpha_{\ \ \sigma \rho \mu} B_\alpha^a (B_a  \cdot B_b) \epsilon^{\mu \rho \sigma \beta} \,, \nonumber \\
\tilde{\mathcal{L}}_{4,2}^4 &\equiv& B_{\beta a} R^\alpha_{\ \ \sigma \rho \mu} B_\alpha^a (B^b  \cdot B_b) \epsilon^{\mu \rho \sigma \beta} \,. \label{L42LPb}
\end{eqnarray}
In these expressions, $g$ is the determinant of the metric, $m_P$ is the reduced Planck mass, $R$ is the Ricci scalar, $B_\mu^a$ is the vector field that belongs to the Lie algebra of the SU(2) group, $A_{\mu \nu}^a \equiv \nabla_\mu B_\nu^a - \nabla_\nu B_\mu^a$ is the Abelian version of the non-Abelian gauge-field strength tensor $F_{\mu \nu}^a$ where $\nabla_\mu$ is the space-time covariant derivative operator, $S_{\mu \nu}^a \equiv \nabla_\mu B_\nu^a + \nabla_\nu B_\mu^a$ is the symmetric version of $A_{\mu \nu}^a$, $G_{\mu \nu}$ is the Einstein tensor, $R^\sigma_{\ \ \nu \rho \mu}$ is the Riemann tensor, $\tilde{A}^{\mu \nu}_a \equiv \frac{1}{2} \epsilon^{\mu \nu \rho \sigma} A_{\rho \sigma a}$ is the Hodge dual of $A^{\mu \nu}_a$, $\epsilon_{abc}$ is the structure-constant tensor of the SU(2) group, $\alpha_{4,0}$, $\alpha_{5,0}$, $\alpha_i$, and $\tilde{\alpha}_i$ are arbitrary dimensionless constants, and $\epsilon^{\mu \nu \rho \sigma}$ is the Levi-Civita tensor of the space-time manifold.  It is worth noting that those Lagrangians with a tilde on top explicitly violate parity while those without a tilde do not.  The exception is $\mathcal{L}_2$ which is an arbitrary function of $A_{\mu \nu}^a$ and $B_\mu^a$, i.e., it can be any scalar built from contractions of the latter two objects with the metric and Levi-Civita tensors of the space-time and SU(2) group manifolds and, therefore, may include both parity-conserving and parity-violating terms. 

Eleven of the Lagrangians in Eqs. \eqn{EH}-\eqn{L50} present decoupling limits that either vanish or produce second-order field equations for the (scalar) longitudinal modes of the vector fields.  The other three Lagrangian pieces present decoupling limits that produce higher-order field equations and, therefore, would give way to the Ostrogradski's instability if they are non degenerate. 

Two of the latter are parity-conserving beyond SU(2) Proca terms and are present in \eq{L42LPa}:
\be \label{bterm1}
\mathcal{L}_{4,2}^3 \equiv B^{\mu a} R^\alpha_{\ \ \sigma \rho \mu} B_{\alpha a} B^{\rho b}  B^\sigma_b + \frac{3}{4} (B_b \cdot B^b) (B^a \cdot B_a) R \,, 
\ee
and
\be \label{bterm2}
\mathcal{L}_{4,2}^4 \equiv [(B_b \cdot B^b) (B^a \cdot B_a) + 2 (B_a \cdot B_b) (B^a \cdot B^b)] R \,.
\ee

The other Lagrangian piece is the purely non-Abelian and parity-violating term which is present in \eq{L42LPb}:
\be\label{l421}
\tilde{\mathcal{L}}_{4,2}^1 \equiv -2 A_{\mu \nu}^a S^{\mu b}_\sigma B_{\alpha a} B_{\beta b} \epsilon^{\nu \sigma \alpha \beta} + S^a_{\mu \nu} S^{\nu b}_\sigma B_{\alpha a} B_{\beta b} \epsilon^{\mu \sigma \alpha \beta} \,.
\ee

We now proceed to write these three Lagrangian pieces in their decoupling limits in order to study the healthiness of their longitudinal modes.

\subsection{Decoupling limit} \label{dl}
The decoupling limit of the GSU2P is found by using the replacement $B^{a}_{\mu}\rightarrow\nabla_{\mu}\phi^{a}$, with $\phi^a$ being a scalar field living in the adjoint representation of the SU(2) group. Nonetheless, to simplify the calculations, instead of using the decoupling limits of the beyond SU(2) Proca Lagrangian pieces in Eqs. \eqn{bterm1} and \eqn{bterm2}, we will use their equivalent forms in the decoupling limit given, respectively, by\footnote{For more details, see Eqs. (52), (56), (62), (63), and (66) of Ref. \cite{GallegoCadavid:2020dho}}:
\be\label{bterm1eq}
\mathcal{L}_{4,2}^3 \Big|_{\rm{sl}} =  \frac{1}{4}[3\mathcal{L}_4^{10}+3\mathcal{L}_4^{11}-4\mathcal{L}_4^{12}-2\mathcal{L}_4^{13}-3\mathcal{L}_4^{14}-3\mathcal{L}_4^{15}+4\mathcal{L}_4^{16}+\mathcal{L}_4^{17}+\mathcal{L}_4^{18}] \,,
\ee
and
\be\label{bterm2eq}
\mathcal{L}_{4,2}^4  \Big|_{\rm{sl}} = -2[-\mathcal{L}_4^{10}-2\mathcal{L}_4^{11}+\mathcal{L}_4^{12}+2\mathcal{L}_4^{13}+\mathcal{L}_4^{14}+2\mathcal{L}_4^{15}-\mathcal{L}_4^{16}-\mathcal{L}_4^{17}-\mathcal{L}_4^{18}] \,,
\ee
where the $\mathcal{L}_4^i$ are given by the corresponding decoupling limits of the following Lagrangians (see Ref. \cite{GallegoCadavid:2020dho}):
\begin{eqnarray}
\mathcal{L}_4^1 &\equiv& A_{\mu \nu}^a S^\mu_{\sigma a} B^{\nu c} B^\sigma_c \,, \nonumber \\
\mathcal{L}_4^2 &\equiv& A_{\mu \nu}^a S^{\mu b}_\sigma B^\nu_a B^\sigma_b \,, \nonumber \\
\mathcal{L}_4^3 &\equiv& A_{\mu \nu}^a S^{\mu b}_\sigma B^\nu_b B^\sigma_a \,, \nonumber \\
\mathcal{L}_4^4 &\equiv& A_{\mu \nu}^a S^{\rho b}_\rho B^\mu_a B^\nu_b \,, \nonumber \\
\mathcal{L}_4^5 &\equiv& A_{\mu \nu}^a S^{\nu b}_\sigma B_{\alpha a} B_{\beta b} \epsilon^{\mu \sigma \alpha \beta} \,, \nonumber \\
\mathcal{L}_4^6 &\equiv& A_{\mu \nu}^a S^{\rho b}_\rho B_{\alpha a} B_{\beta b} \epsilon^{\mu \nu \alpha \beta} \,, \nonumber \\
\mathcal{L}_4^7 &\equiv& A_{\mu \nu}^a S_{\rho \sigma a} B^{\rho c} B_{\beta c} \epsilon^{\mu \nu \sigma \beta} \,, \nonumber \\
\mathcal{L}_4^8 &\equiv& A_{\mu \nu}^a S^b_{\rho \sigma} B^\rho_a B_{\beta b} \epsilon^{\mu \nu \sigma \beta} \,, \nonumber \\
\mathcal{L}_4^9 &\equiv& A_{\mu \nu}^a S^b_{\rho \sigma} B^\rho_b B_{\beta a} \epsilon^{\mu \nu \sigma \beta} \,, \nonumber \\
\mathcal{L}_4^{10} &\equiv& S^{\mu a}_\mu S^\rho_{\rho a} (B^c \cdot B_c) \,, \nonumber \\
\mathcal{L}_4^{11} &\equiv& S^{\mu a}_\mu S^{\rho b}_\rho (B_a \cdot B_b) \,, \nonumber \\
\mathcal{L}_4^{12} &\equiv& S^{\mu a}_\mu S_{\rho \sigma a} B^{\rho c} B^\sigma_c \,, \nonumber \\
\mathcal{L}_4^{13} &\equiv& S^{\mu a}_\mu S^b_{\rho \sigma} B^\rho_a B^\sigma_b \,, \nonumber \\
\mathcal{L}_4^{14} &\equiv& S^a_{\mu \nu} S^{\mu \nu}_a (B^c \cdot B_c) \,, \nonumber \\
\mathcal{L}_4^{15} &\equiv& S^a_{\mu \nu} S^{\mu \nu b} (B_a \cdot B_b) \,, \nonumber \\
\mathcal{L}_4^{16} &\equiv& S^a_{\mu \nu} S^\mu_{\sigma a} B^{\nu c} B^\sigma_c \,, \nonumber \\
\mathcal{L}_4^{17} &\equiv& S^a_{\mu \nu} S^{\mu b}_\sigma B^\nu_a B^\sigma_b \,, \nonumber \\
\mathcal{L}_4^{18} &\equiv& S^a_{\mu \nu} S^{\mu b}_\sigma B^\nu_b B^\sigma_a \,, \nonumber \\
\mathcal{L}_4^{19} &\equiv& S^a_{\mu \nu} S^{\nu b}_\sigma B_{\alpha a} B_{\beta b} \epsilon^{\mu \sigma \alpha \beta} \,.
\end{eqnarray}
It is worth mentioning that, in Eqs. \eqn{bterm1eq} and \eqn{bterm2eq}, we have neglected terms proportional to the Einstein tensor $G_{\mu \nu}$, to $\mathcal{L}_{4,2}^2$ in \eq{L42LPa}, and to 
\begin{equation}
\hat{\mathcal{L}}_4^{1,h} \equiv \frac{1}{4} (B_b \cdot B^b) [S^{\mu a}_\mu S^\nu_{\nu a} - S^{\mu a}_\nu S^\nu_{\mu a} - R (B^a \cdot B_a)] + \frac{1}{2} (B_a \cdot B_b) [S^{\mu a}_\mu S^{\nu b}_\nu - S^{\mu a}_\nu S^{\nu b}_\mu - R (B^a \cdot B^b)] \,, \nonumber 
\end{equation}
(see Eq. (57) of Ref. \cite{GallegoCadavid:2020dho}), since they are
healthy in the decoupling limit. 

Thus, in the decoupling limit, the beyond SU(2) Proca terms in Eqs. \eqn{bterm1eq} and \eqn{bterm2eq} can be written, respectively, as
\begin{align}\label{bterm1eqsl}
 \mathcal{L}_{4,2}^3 \Big|_{\rm{sl}} &=3\left(\nabla_{\alpha}\phi^{b}\right)\left(\nabla^{\alpha}\phi^{a}\right)\left(\nabla_{\mu}\nabla^{\mu}\phi_{a}\right)\left(\nabla_{\nu}\nabla^{\nu}\phi_{b}\right)+3\left(\nabla_{\alpha}\phi_{a}\right) \left(\nabla^{\alpha}\phi^{a}\right) \left(\nabla_{\mu}\nabla^{\mu}\phi^{b}\right)\left(\nabla_{\nu}\nabla^{\nu}\phi_{b}\right) \notag \\
 & -4\left(\nabla^{\alpha}\phi^{a}\right)\left(\nabla_{\mu}\nabla_{\alpha}\phi^{b}\right)\left(\nabla^{\mu}\phi_{a}\right)\left(\nabla_{\nu}\nabla^{\nu}\phi_{b}\right) -2\left(\nabla^{\alpha}\phi^{a}\right)\left(\nabla_{\mu}\nabla_{\alpha}\phi_{a}\right)\left(\nabla^{\mu}\phi^{b}\right)\left(\nabla_{\nu}\nabla^{\nu}\phi_{b}\right) \notag \\
 &+\left(\nabla^{\alpha}\phi^{a}\right)\left(\nabla^{\mu}\phi^{b}\right)\left(\nabla_{\nu}\nabla_{\mu}\phi_{b}\right)\left(\nabla^{\nu}\nabla_{\alpha}\phi_{a}\right)+4\left(\nabla^{\alpha}\phi^{a}\right)\left(\nabla^{\mu}\phi_{a}\right)\left(\nabla_{\nu}\nabla_{\mu}\phi_{b}\right)\left(\nabla^{\nu}\nabla_{\alpha}\phi^{b}\right) \\
 & +\left(\nabla^{\alpha}\phi^{a}\right)\left(\nabla^{\mu}\phi^{b}\right)\left(\nabla_{\nu}\nabla_{\alpha}\phi_{b}\right)\left(\nabla^{\nu}\nabla_{\mu}\phi_{a}\right)-3\left(\nabla_{\alpha}\phi^{b}\right)\left(\nabla^{\alpha}\phi^{a}\right)\left(\nabla_{\nu}\nabla_{\mu}\phi_{b}\right)\left(\nabla^{\nu}\nabla^{\mu}\phi_{a}\right) \notag \\
 & -3\left(\nabla_{\alpha}\phi_{a}\right)\left(\nabla^{\alpha}\phi^{a}\right)\left(\nabla_{\nu}\nabla_{\mu}\phi_{b}\right)\left(\nabla^{\nu}\nabla^{\mu}\phi^{b}\right) \notag \,,
  \end{align}
 and 
 \begin{align}\label{bterm2eqsl}
 \mathcal{L}_{4,2}^4 \Big|_{\rm{sl}} &= 16\left(\nabla_{\alpha}\phi^{b}\right)\left(\nabla^{\alpha}\phi^{a}\right)\left(\nabla_{\mu}\nabla^{\mu}\phi_{a}\right)\left(\nabla_{\nu}\nabla^{\nu}\phi_{b}\right)+8\left(\nabla_{\alpha}\phi_{a}\right)\left(\nabla^{\alpha}\phi^{a}\right)\left(\nabla_{\mu}\nabla^{\mu}\phi^{b}\right)\left(\nabla_{\nu}\nabla^{\nu}\phi_{b}\right)\notag \\
 & -8\left(\nabla^{\alpha}\phi^{a}\right)\left(\nabla_{\mu}\nabla_{\alpha}\phi^{b}\right)\left(\nabla^{\mu}\phi_{a}\right)\left(\nabla_{\nu}\nabla^{\nu}\phi_{b}\right)-16\left(\nabla^{\alpha}\phi^{a}\right)\left(\nabla_{\mu}\nabla_{\alpha}\phi_{a}\right)\left(\nabla^{\mu}\phi^{b}\right)\left(\nabla_{\nu}\nabla^{\nu}\phi_{b}\right) \notag\\
 & +8\left(\nabla^{\alpha}\phi^{a}\right)\left(\nabla^{\mu}\phi^{b}\right)\left(\nabla_{\nu}\nabla_{\mu}\phi_{b}\right)\left(\nabla^{\nu}\nabla_{\alpha}\phi_{a}\right)+8\left(\nabla^{\alpha}\phi^{a}\right)\left(\nabla^{\mu}\phi_{a}\right)\left(\nabla_{\nu}\nabla_{\mu}\phi_{b}\right)\left(\nabla^{\nu}\nabla_{\alpha}\phi^{b}\right)  \\
 &+8\left(\nabla^{\alpha}\phi^{a}\right)\left(\nabla^{\mu}\phi^{b}\right)\left(\nabla_{\nu}\nabla_{\alpha}\phi_{b}\right)\left(\nabla^{\nu}\nabla_{\mu}\phi_{a}\right)-16\left(\nabla_{\alpha}\phi^{b}\right)\left(\nabla^{\alpha}\phi^{a}\right)\left(\nabla_{\nu}\nabla_{\mu}\phi_{b}\right)\left(\nabla^{\nu}\nabla^{\mu}\phi_{a}\right) \notag \\
 &-8\left(\nabla_{\alpha}\phi_{a}\right)\left(\nabla^{\alpha}\phi^{a}\right)\left(\nabla_{\nu}\nabla_{\mu}\phi_{b}\right)\left(\nabla^{\nu}\nabla^{\mu}\phi^{b}\right) \,. \notag
 \end{align}

For the term $\tilde{\mathcal{L}}_{4,2}^1$, we have the following expression for the decoupling limit:
\begin{align}\label{l421sl}
 \tilde{\mathcal{L}}_{4,2}^1 \Big|_{\rm{sl}} =4\epsilon_{\alpha\beta\rho\sigma}\left(\nabla^{\alpha}\phi^{a}\right)\left(\nabla^{\beta}\phi^{b}\right)\left(\nabla^{\rho}\nabla^{\lambda}\phi_{a}\right)\left(\nabla^{\sigma}\nabla_{\lambda}\phi_{b}\right) \,.
 \end{align}

The purpose of the following sections is to study the degeneracy of the kinetic matrices associated to the decoupling-limit Lagrangians shown in Eqs.~\eqn{bterm1eqsl}-\eqn{l421sl}. To do this, we follow the procedure described in Ref. \cite{GallegoCadavid:2021ljh} employing the $3+1$ ADM formalism \cite{3+1book}.

\section{3+1 decomposition} \label{decomp}
To study the degeneracy properties of a general Lagrangian containing an SU(2) vector field $B_{\mu}^a$ in the decoupling limit, we introduce an auxiliary field $Z^{a}_{\mu}$ which leads to an action with first-order derivatives only\footnote{To avoid confusion, we follow the notation of Ref. \cite{GallegoCadavid:2021ljh} in the decoupling-limit case throughout this paper.}. Therefore, we write $Z^{a}_{\mu}\equiv\nabla_{\mu}\phi^{a}$ not to confuse the auxiliary field with the original field $B^{a}_{\mu}$. Then, we split the time derivatives from the spatial ones using the covariant 3+1 decomposition of the spacetime~\cite{Langlois:2015cwa, Kimura:2016rzw,GallegoCadavid:2021ljh}.
To do so, we introduce a time-like unit vector $n^\mu$ normal to the spatial hypersurfaces $\Sigma_{t}$. This foliation induces a three-dimensional metric $h_{\mu\nu} \equiv g_{\mu\nu} + n_{\mu} n_{\nu}$ on the spatial hypersurfaces $\Sigma_{t}$ \cite{Langlois:2015cwa,Kimura:2016rzw,GallegoCadavid:2021ljh,3+1book}. Thus, the SU(2) field $Z_{\mu}^a$ can be decomposed as
\begin{equation}\label{adecom}
Z^{a}_{\mu}=-Z^{a}_{*}n_{\mu}+\hat{Z}^{a}_{\mu} \,,
\end{equation}
where $Z_{*}^a \equiv n^{\mu}Z_{\mu}^a$ and $\hat{Z}_{\mu}^a \equiv h_{\mu}^{\nu}Z_{\nu}^a$ are the normal and spatial projections of $Z_\mu^a$, respectively. The ``time derivative'' of any  spatial tensor, denoted by a dot, is defined as the spatial projection of its Lie derivative  with respect to the time direction vector $t^\mu \equiv \partial/\partial t$, the latter being associated with a time coordinate $t$ that labels the slicing of space-like hypersurfaces. 

Now, using the above definitions, a kinetic Lagrangian consisting of SU(2) vectors in the decoupling limit can be written in terms of $\dot{Z}^{a}_{*}$ and the extrinsic curvature $K_{\mu\nu}$ only as \cite{GallegoCadavid:2021ljh} \begin{equation}
   \mathcal{L}_{\rm kin}=\tilde{\mathcal{A}}_{ab}\dot{Z}^{a}_{*}\dot{Z}^{b}_{*}
  +\tilde{\mathcal{F}}^{\alpha\beta\rho\sigma}K_{\alpha\beta}K_{\rho\sigma} +2\tilde{\mathcal{C}}^{\alpha\beta}_{a}\dot{Z}^{a}_{*}K_{\alpha\beta} \,.
\end{equation}
This kinetic Lagrangian does not depend on $\dot{\hat{Z}}^{a}_{\mu}$, which can be removed via the relation $\nabla_{\mu}Z^{a}_{\nu}=\nabla_{\nu}Z^{a}_{\mu}$ \cite{Langlois:2015cwa, Kimura:2016rzw, GallegoCadavid:2021ljh}. In general, this action can be written in a matrix form as follows:
\begin{equation}
 \mathcal{L}_{\rm kin}=
  \begin{bmatrix}
\dot{Z}^{a}_{*} & K_{\rho\sigma} 
    \end{bmatrix}
    \begin{bmatrix}
\tilde{\mathcal{A}}_{ab} & \tilde{\mathcal{C}}^{\alpha\beta}_{a} \\	
\tilde{\mathcal{C}}^{\rho\sigma}_{b} & \tilde{\mathcal{F}}^{\alpha\beta\rho\sigma}
    \end{bmatrix} 
        \begin{bmatrix}
\dot{Z}^{b}_{*}  \\
K_{\alpha\beta} 
    \end{bmatrix} \,.
    \label{eq:kinm}
\end{equation}

Our first task is to calculate the determinant of the kinetic matrix given in \eq{eq:kinm}. We will diagonalize the three resulting matrices in \eq{eq:kinm}, at least partially, by performing two basis transformations. In the next section, with the block-diagonal matrix, it will be easier to calculate the determinant for each of the corresponding Lagrangians of interest.

\subsection{Change of basis}\label{cb}
Following the methodology outlined in Ref. \cite{GallegoCadavid:2021ljh}, we introduce a pair of basis vectors: one for the internal SU(2) space and another one for the spatial hypersurface $\Sigma_t$. For the internal space, we have the set of basis vectors
\begin{equation}
\{W^{a}_{i}\}=\left\{W^{a}_{1}=\frac{Z^{a}_{*}}{|Z^{a}_{*}|}, W^{a}_{2}, W^{a}_{3}\right\} \ \mbox{fulfilling } \ W^{a}_{i}W^{j}_{a}=\delta^{j}_{i}\,.
\label{eq:basisgauge}
\end{equation}
Notice that the $i$-index labels the vectors in the basis. On the hypersurfaces, we define three basis vectors $V_\mu^i$ orthogonal to  $n^\mu$:
\begin{equation}
\{V^{i}_{\mu}\}=\left\{V^{1}_{\mu}=\frac{\hat{Z}^{1}_{\mu}}{|\hat{Z}^{1}_{\mu}|},V^{2}_{\mu}, V^{3}_{\mu}\right\} \ \mbox{fulfilling } \ V^{i}_{\mu}V^{\mu}_{j}=\delta^{i}_{j}, \ V^{i}_{\mu}n^{\mu}=0 \,,
\label{eq:basishyper}
\end{equation}
where again the $i$-index in the $V_\mu^i$ corresponds to the labeling of the three basis vectors.

Thus, using the two bases in Eqs. (\ref{eq:basisgauge}) and (\ref{eq:basishyper}), we can now decompose the vector field $Z_\mu^a$ as 
\begin{equation}
    Z_{*}^{a}= \tilde{Z_{*}}^{i} W^{a}_{i} \, \quad \mbox{and} \quad \hat{Z}^{a}_{\mu}= \tilde{\hat{Z}}^{k}_{i} W^{a}_{k}V^{i}_{\mu} \,,
\end{equation}
where $i$, $j$ and $k$ run from 1 to 3. Following Ref.~\cite{GallegoCadavid:2021ljh}, from now on, we set $\tilde{Z_{*}}^{i}=\tilde{\hat{Z}}^{k}_{i}=0$ for $i \neq 1$.

Finally, we define six independent symmetric matrices ($U^I_{\mu\nu}, \ I=1,...,6$): 
\bea
U^1_{\mu\nu} \equiv V^1_\mu V^1_\nu \,, \quad 
U^2_{\mu\nu} &\equiv& \frac{1}{\sqrt{2}} (h_{\mu\nu} - U^1_{\mu\nu}) \,, \quad \quad \quad
U^3_{\mu\nu} \equiv \frac{1}{\sqrt{2}}(V^2_\mu V^2_\nu - V^3_\mu V^3_\nu) \,, \notag\\
U^4_{\mu\nu} \equiv \frac{1}{\sqrt{2}}(V^2_\mu V^3_\nu + V^2_\nu V^3_\mu) \,,  \quad 
U^5_{\mu\nu} &\equiv& \frac{1}{ \sqrt{2} }(V^2_\mu V^1_\nu + V^2_\nu V^1_\mu) \,, \quad 
U^6_{\mu\nu} \equiv \frac{1}{\sqrt{2 } }(V^3_\mu V^1_\nu + V^3_\nu V^1_\mu) \,,
\eea
such that  $K_{\mu\nu}$ can be decomposed as $K_{\mu\nu} = K_I \, U_{\mu\nu}^I$ \cite{GallegoCadavid:2021ljh}.

With all the previous definitions, we can rewrite the 
kinetic Lagrangian in \eq{eq:kinm} as
\bea
\mathcal{L}_{\rm kin}=
\left(
\begin{array}{cc}
	\tilde{{\bm u}}_{1}^T & \tilde{{\bm u}}_{2}^T \\
\end{array}
\right)
\left(
\begin{array}{cc}
	\tilde{\mathcal{M}}_{1} & 0  \\
	0 & \tilde{\mathcal{M}}_{2}  \\
\end{array}
\right)
\left(
\begin{array}{cc}
	\tilde{{\bm u}}_{1} \\
	{\bm \tilde{{\bm u}}_{2}}  \\
\end{array}
\right)
,
\label{Lkins}
\eea
where the vector components are
$\tilde{u}_{1} \equiv (\dot{\tilde{Z}}^{1}_{*},\dot{\tilde{Z}}^{2}_{*},\dot{\tilde{Z}}^{3}_{*},K_{1},K_{2})$ 
and $\tilde{u}_{2} \equiv (K_{5},K_{6}, K_{3},K_{4})$. 

Hence, by performing the outlined change of basis, we have been able to partially diagonalize the kinetic matrix. In this form, it will be easier to calculate its determinant.

\section{Determinant of the kinetic matrices}\label{det}
We calculate in this section the kinetic matrix determinant for each of the three relevant Lagrangian pieces discussed in the introduction. In the process, we show explicitly the entries of the kinetic matrices given by Eqs. \eqn{eq:kinm} and \eqn{Lkins} corresponding to the matrices before and after the change of basis, respectively.

\subsection{Kinetic Lagrangian for $\mathcal{L}_{4,2}^3 \Big|_{\rm{sl}}$}
Let us begin with the Lagrangian piece $\mathcal{L}_{4,2}^3 \Big|_{\rm{sl}}$. In this case, the elements in the matrix of \eq{eq:kinm} are given by
\begin{equation}
    \tilde{\mathcal{A}}_{ab}\equiv 0 \,,
\end{equation}
\begin{align}
    \tilde{\mathcal{C}}^{\rho\sigma}_{a}\equiv -4\hat{Z}^{b\rho}\hat{Z}^{\sigma}_{b}Z_{*a}-\left(\hat{Z}^{b\rho}\hat{Z}^{\sigma}_{a}+\hat{Z}^{\rho}_{a}\hat{Z}^{b\sigma}\right)Z_{*b}+6\left(\hat{Z}^{2}-Z_{*}^{2}\right)Z_{*a}h^{\rho\sigma}+6\hat{Z}_{\alpha}^{b}\hat{Z}^{\alpha}_{a}Z_{*b}h^{\rho\sigma} \,,
\end{align}
and
\begin{align}
    \tilde{\mathcal{F}}^{\alpha\beta\rho\sigma}\equiv& -\frac{5}{2}\hat{Z}^{\alpha}_{c}\hat{Z}^{\beta}_{b}\hat{Z}^{b\rho}\hat{Z}^{c\sigma} -\frac{5}{2}\hat{Z}^{\alpha}_{c}\hat{Z}^{\beta}_{b}\hat{Z}^{c\rho}\hat{Z}^{b\sigma}-\hat{Z}^{\alpha}_{c}\hat{Z}^{c\beta}\hat{Z}^{b\rho}\hat{Z}^{\sigma}_{b}-Z_{*}^{2}\hat{Z}^{b\rho}\hat{Z}^{\sigma}_{b}h^{\alpha\beta}\notag \\
    & +4\hat{Z}^{\rho}_{b}\hat{Z}^{\sigma}_{c}Z_{*}^{b}Z_{*}^{c}h^{\alpha\beta}+\frac{3}{2}\hat{Z}^{2}\hat{Z}^{\beta}_{c}\hat{Z}^{c\sigma}h^{\alpha\rho}+\frac{3}{2}\hat{Z}^{\lambda}_{b}\hat{Z}^{c}_{\lambda}\hat{Z}^{b\beta}\hat{Z}^{\sigma}_{c}h^{\alpha\rho}-3\hat{Z}^{\beta}_{b}\hat{Z}^{\sigma}_{c}Z_{*}^{b}Z_{*}^{c}h^{\alpha\rho} \notag \\
    & +\frac{3}{2}\hat{Z}^{2}\hat{Z}^{\beta}_{c}\hat{Z}^{c\rho}h^{\alpha\sigma}+\frac{3}{2}\hat{Z}^{\lambda}_{b}\hat{Z}^{c}_{\lambda}\hat{Z}^{\beta}_{c}\hat{Z}^{b\rho}h^{\alpha\sigma}-3\hat{Z}^{\beta}_{c}\hat{Z}^{\rho}_{b}Z_{*}^{c}Z_{*}^{b}h^{\alpha\sigma}+\frac{3}{2}\hat{Z}^{2}\hat{Z}^{\alpha}_{c}\hat{Z}^{c\sigma}h^{\beta\rho} \notag \\
    &+\frac{3}{2}\hat{Z}^{\lambda}_{b}\hat{Z}_{\lambda}^{c}\hat{Z}^{\alpha}_c\hat{Z}^{b\sigma}h^{\beta\rho}-3\hat{Z}^{\alpha}_{c}\hat{Z}^{\sigma}_{b}Z_{*}^{c}Z_{*}^{b}h^{\beta\rho}-\frac{3}{2}\hat{Z}^{2}Z_{*}^{2}h^{\alpha\sigma}h^{\beta\rho}+3Z_{*}^{4}h^{\alpha\sigma}h^{\beta\rho} \notag \\
    &-\frac{3}{2}\hat{Z}^{\lambda}_{c}\hat{Z}_{b\lambda}Z_{*}^{c}Z_{*}^{b}h^{\alpha\sigma}h^{\beta\rho}+\frac{3}{2}\hat{Z}^{2}\hat{Z}^{\alpha}_{c}\hat{Z}^{c\rho}h^{\beta\sigma}+\frac{3}{2}\hat{Z}^{c}_{\lambda}\hat{Z}^{b\lambda}\hat{Z}^{\alpha}_{c}\hat{Z}^{\rho}_{b}h^{\beta\sigma}-3\hat{Z}^{\alpha}_{c}\hat{Z}^{\rho}_{b}Z_{*}^{b}Z_{*}^{c}h^{\beta\sigma} \notag \\
    & -\frac{3}{2}\hat{Z}^{2}Z_{*}^{2}h^{\alpha\rho}h^{\beta\sigma}+3Z_{*}^{4}h^{\alpha\rho}h^{\beta\sigma}-\frac{3}{2}\hat{Z}_{c\lambda}\hat{Z}_{b}^{\lambda}Z_{*}^{b}Z_{*}^{c}h^{\alpha\rho}h^{\beta\sigma} -Z_{*}^{2}\hat{Z}^{\alpha}_{c}\hat{Z}^{c\beta}h^{\rho\sigma} \notag \\
    & +4\hat{Z}^{\alpha}_{c}\hat{Z}^{\beta}_{b}Z_{*}^{c}Z_{*}^{b}h^{\rho\sigma}+3\hat{Z}^{2}Z_{*}^{2}h^{\alpha\beta}h^{\rho\sigma}-6Z_{*}^{4}h^{\alpha\beta}h^{\rho\sigma}+3\hat{Z}_{b\lambda}\hat{Z}^{\lambda}_{c}Z_{*}^{c}Z_{*}^{b}h^{\alpha\beta}h^{\rho\sigma} \,.
\end{align}

Now, after the change of basis in Sec. \ref{cb}, we get the structure given in \eq{Lkins} with the submatrices
\bea \label{m1l1}
\tilde{\mathcal{M}}_{1}=
\left(
\begin{array}{ccccc}
0 & 0 & 0 & \tilde{\mathcal{C}}_{1} & \tilde{\mathcal{C}}_{2} \\
0 & 0 & 0 & \tilde{\mathcal{C}}_{3} & \tilde{\mathcal{C}}_{4} \\
0 & 0 & 0 & \tilde{\mathcal{C}}_{5} & \tilde{\mathcal{C}}_{6} \\
\tilde{\mathcal{C}}_{1} & \tilde{\mathcal{C}}_{3} & \tilde{\mathcal{C}}_{5} & \tilde{\mathcal{F}}_{1} & \tilde{\mathcal{F}}_{2} \\
\tilde{\mathcal{C}}_{2} & \tilde{\mathcal{C}}_{4} & \tilde{\mathcal{C}}_{6} & \tilde{\mathcal{F}}_{2} & \tilde{\mathcal{F}}_{3} \\
\end{array}
\right)
,
\eea
and
\bea \label{m2l1}
\tilde{\mathcal{M}}_{2}=
\left(
\begin{array}{cccc}
	 \tilde{\mathcal{F}}_{4}  & 0 & 0 & 0 \\
	0 &  \tilde{\mathcal{F}}_{4}  & 0 & 0 \\
    0 & 0 &  \tilde{\mathcal{F}}_{5}  & 0 \\
    0 & 0 & 0 &  \tilde{\mathcal{F}}_{5}  \\
\end{array}
\right)
,
\eea
where each entry of these submatrices is given in the Appendix \ref{aA}.

From the form of $\tilde{\mathcal{M}}_{1}$, it is easy to check that its determinant is zero.  Thus, the $\mathcal{L}_{4,2}^3\Big|_{\rm{sl}}$ Lagrangian is degenerate.

\subsection{Kinetic Lagrangian for $\mathcal{L}_{4,2}^4 \Big|_{\rm{sl}}$}
In a similar way, the $\mathcal{L}_{4,2}^4 \Big|_{\rm{sl}}$ Lagrangian generates the following kinetic matrix elements:
\begin{equation}
    \tilde{\mathcal{A}}_{ab}\equiv0 \,,
\end{equation}
\begin{align}
    \tilde{\mathcal{C}}^{\rho\sigma}_{a}\equiv-8\hat{Z}^{b\rho}\hat{Z}^{\sigma}_{b}Z_{*a}-8\left(\hat{Z}^{b\rho}\hat{Z}^{\sigma}_{a}+\hat{Z}^{\rho}_{a}\hat{Z}^{b\sigma}\right)Z_{*b}+\left(16\hat{Z}^{2}-24Z_{*}^{2}\right)Z_{*a}h^{\rho\sigma}+32\hat{Z}_{\alpha}^{b}\hat{Z}^{\alpha}_{a}Z_{*b}h^{\rho\sigma} \,,
\end{align}
and
\begin{align}
    \tilde{\mathcal{F}}^{\alpha\beta\rho\sigma}&\equiv-8\hat{Z}^{\alpha}_{c}\hat{Z}^{\beta}_{b}\hat{Z}^{b\rho}\hat{Z}^{c\sigma}-8\hat{Z}^{\alpha}_{c}\hat{Z}^{\beta}_{b}\hat{Z}^{c\rho}\hat{Z}^{b\sigma}-8\hat{Z}^{\alpha}_{c}\hat{Z}^{c\beta}\hat{Z}^{\rho}_{b}\hat{Z}^{b\sigma}+4Z_{*}^{2}\hat{Z}^{\rho}_{c}\hat{Z}^{c\sigma}h^{\alpha\beta} \notag \\
    & +8\hat{Z}^{\rho}_{c}\hat{Z}^{\sigma}_{b}Z_{*}^{b}Z_{*}^{c}h^{\alpha\beta}+4\left(\hat{Z}^{2}-Z_{*}^{2}\right)\hat{Z}^{\beta}_{c}\hat{Z}^{c\sigma}h^{\alpha\rho}+8\hat{Z}^{\lambda}_{c}\hat{Z}_{b\lambda}\hat{Z}^{c\beta}\hat{Z}^{b\sigma}h^{\alpha\rho}-8\hat{Z}^{\beta}_{c}\hat{Z}^{\sigma}_{b}Z_{*}^{c}Z_{*}^{b}h^{\alpha\rho} \notag \\
    &+4\left(\hat{Z}^{2}-Z_{*}^{2}\right)\hat{Z}^{\beta}_{c}\hat{Z}^{c\rho}h^{\alpha\sigma}+8\hat{Z}^{c\lambda}\hat{Z}^{b}_{\lambda}\hat{Z}^{\beta}_{c}\hat{Z}^{\rho}_{b}h^{\alpha\sigma}-8\hat{Z}^{\beta}_{c}\hat{Z}^{\rho}_{b}Z_{*}^{b}Z_{*}^{c}h^{\alpha\sigma}+4\left(\hat{Z}^{2}-Z_{*}^{2}\right)\hat{Z}^{\alpha}_{c}\hat{Z}^{c\sigma}h^{\beta\rho} \notag \\
    & +8\hat{Z}^{c}_{\lambda}\hat{Z}^{b\lambda}\hat{Z}^{\alpha}_{c}\hat{Z}^{\sigma}_{b}h^{\beta\rho}-8\hat{Z}^{\alpha}_{c}\hat{Z}^{\sigma}_{b}Z_{*}^{b}Z_{*}^{c}h^{\beta\rho}-4\hat{Z}^{2}Z_{*}^{2}h^{\alpha\sigma}h^{\beta\rho}+12Z_{*}^{4}h^{\alpha\sigma}h^{\beta\rho}-8\hat{Z}^{\lambda}_{c}\hat{Z}_{b\lambda}Z_{*}^{c}Z_{*}^{b}h^{\alpha\sigma}h^{\beta\rho} \notag \\
    & +4\left(\hat{Z}^{2}-Z_{*}^{2}\right)\hat{Z}^{\alpha}_{c}\hat{Z}^{c\rho}h^{\beta\sigma}+8\hat{Z}^{c\lambda}\hat{Z}_{\lambda}^{b}\hat{Z}^{\alpha}_{c}\hat{Z}^{\rho}_{b}h^{\beta\sigma}-8\hat{Z}^{\alpha}_{c}\hat{Z}^{\rho}_{b}Z_{*}^{c}Z_{*}^{b}h^{\beta\sigma} -4\hat{Z}^{2}Z_{*}^{2}h^{\alpha\rho}h^{\beta\sigma}+12Z_{*}^{4}h^{\alpha\rho}h^{\beta\sigma} \notag \\
    &-8\hat{Z}_{c\lambda}\hat{Z}_{b}^{\lambda}Z_{*}^{c}Z_{*}^{b}h^{\alpha\rho}h^{\beta\sigma}+4Z_{*}^{2}\hat{Z}^{\alpha}_{c}\hat{Z}^{c\beta}h^{\rho\sigma}+8\hat{Z}^{\alpha}_{c}\hat{Z}^{\beta}_{b}Z_{*}^{c}Z_{*}^{b}h^{\rho\sigma}+8\hat{Z}^{2}Z_{*}^{2}h^{\alpha\beta}h^{\rho\sigma}-24Z_{*}^{4}h^{\alpha\beta}h^{\rho\sigma} \notag \\
    & +16\hat{Z}^{\lambda}_{c}\hat{Z}_{b\lambda}Z_{*}^{c}Z_{*}^{b}h^{\alpha\beta}h^{\rho\sigma} \,.
\end{align}

Now, after the change of basis in Sec. \ref{cb}, we get the structure given in \eq{Lkins} with the submatrices

\bea\label{m1l2}
\tilde{\mathcal{M}}_{1}=
\left(
\begin{array}{ccccc}
0 & 0 & 0 & \tilde{\mathcal{C}}_{7} & \tilde{\mathcal{C}}_{8} \\
0 & 0 & 0 & \tilde{\mathcal{C}}_{9} & \tilde{\mathcal{C}}_{10} \\
0 & 0 & 0 & \tilde{\mathcal{C}}_{11} & \tilde{\mathcal{C}}_{12} \\
\tilde{\mathcal{C}}_{7} & \tilde{\mathcal{C}}_{9} & \tilde{\mathcal{C}}_{11} & \tilde{\mathcal{F}}_{6} & \tilde{\mathcal{F}}_{7} \\
\tilde{\mathcal{C}}_{8} & \tilde{\mathcal{C}}_{10} & \tilde{\mathcal{C}}_{12} & \tilde{\mathcal{F}}_{7} & \tilde{\mathcal{F}}_{8} \\
\end{array}
\right)
,
\eea
and
\bea\label{m2l2}
\tilde{\mathcal{M}}_{2}=
\left(
\begin{array}{cccc}
	 \tilde{\mathcal{F}}_{9}  & 0 & 0 & 0 \\
	0 &  \tilde{\mathcal{F}}_{9}  & 0 & 0 \\
    0 & 0 &  \tilde{\mathcal{F}}_{10}  & 0 \\
    0 & 0 & 0 &  \tilde{\mathcal{F}}_{10}  \\
\end{array}
\right)
.
\eea
where each entry of these submatrices is given in Appendix \ref{aB}.

From the form of $\tilde{\mathcal{M}}_{1} $, we conclude again that $\det{\tilde{\mathcal{M}}_{1}}=0$ and, hence, the $\mathcal{L}_{4,2}^4\Big|_{\rm{sl}}$ Lagrangian is also degenerate.
\subsection{Kinetic Lagrangian for $\tilde{\mathcal{L}}_{4,2}^1 \Big|_{\rm{sl}}$}
In this case, we get
\begin{equation}
    \tilde{\mathcal{A}}_{ab}\equiv 0 \,,
\end{equation}
\begin{align}
    \tilde{\mathcal{C}}^{\rho\sigma}_{a}\equiv -4\epsilon_{\alpha\beta\lambda\gamma}\hat{Z}^{\alpha}_{a}\hat{Z}^{\beta}_{b}\hat{Z}^{b\sigma}h^{\rho\gamma}n^{\lambda}-4\epsilon_{\alpha\beta\lambda\gamma}\hat{Z}^{\alpha}_{a}\hat{Z}^{\beta}_{b}\hat{Z}^{b\rho}h^{\sigma\gamma}n^{\lambda} \,,
\end{align}
and
\begin{align}
    \tilde{\mathcal{F}}^{\alpha\beta\rho\sigma}&\equiv-\epsilon_{\mu\nu\lambda\gamma}\hat{Z}^{\mu}_{c}\hat{Z}^{\nu}_{b}\hat{Z}^{c\beta}\hat{Z}^{b\sigma}h^{\alpha\lambda}h^{\rho\gamma}-\epsilon_{\mu\nu\lambda\gamma}\hat{Z}^{\mu}_{c}\hat{Z}^{\nu}_{b}\hat{Z}^{c\alpha}\hat{Z}^{b\sigma}h^{\beta\lambda}h^{\rho\gamma}-\epsilon_{\mu\nu\lambda\gamma}\hat{Z}^{\mu}_{c}\hat{Z}^{\nu}_{b}\hat{Z}^{c\beta}\hat{Z}^{b\rho}h^{\alpha\lambda}h^{\sigma\gamma}\notag \\
    & -\epsilon_{\mu\nu\lambda\gamma}\hat{Z}^{\mu}_{c}\hat{Z}^{\nu}_{b}\hat{Z}^{c\alpha}\hat{Z}^{b\rho}h^{\beta\lambda}h^{\sigma\gamma}-\epsilon_{\mu\nu\lambda\gamma}\hat{Z}^{\mu}_{b}\hat{Z}^{b\sigma}\hat{Z}^{\beta}_{c}Z_{*}^{c}h^{\alpha\lambda}h^{\rho\gamma}n^{\nu}+\epsilon_{\mu\nu\lambda\gamma}\hat{Z}^{\mu}_{c}\hat{Z}^{c\beta}\hat{Z}^{\sigma}_{b}Z_{*}^{b}h^{\alpha\lambda}h^{\rho\gamma}n^{\nu} \notag\\
    &-\epsilon_{\mu\nu\lambda\gamma}\hat{Z}^{\mu}_{b}\hat{Z}^{b\sigma}\hat{Z}^{\alpha}_{b}Z_{*}^{b}h^{\beta\lambda}h^{\rho\gamma}n^{\nu} +\epsilon_{\mu\nu\lambda\gamma}\hat{Z}^{\mu}_{c}\hat{Z}^{c\alpha}\hat{Z}^{\sigma}_{b}h^{\beta\lambda}h^{\rho\gamma}n^{\nu}-\epsilon_{\mu\nu\lambda\gamma}\hat{Z}^{\mu}_{b}\hat{Z}^{b\rho}\hat{Z}^{\beta}_{c}Z_{*}^{c}h^{\alpha\lambda}h^{\sigma\gamma}n^{\nu} \notag \\
    & +\epsilon_{\mu\nu\lambda\gamma}\hat{Z}^{\mu}_{c}\hat{Z}^{c\beta}\hat{Z}^{\rho}_{b}Z_{*}^{b}h^{\alpha\lambda}h^{\sigma\gamma}n^{\nu}-\epsilon_{\mu\nu\lambda\gamma}\hat{Z}^{\mu}_{b}\hat{Z}^{b\rho}\hat{Z}^{\alpha}_{c}Z_{*}^{c}h^{\beta\lambda}h^{\sigma\gamma}n^{\nu}+\epsilon_{\mu\nu\lambda\gamma}\hat{Z}^{\mu}_{c}\hat{Z}^{c\alpha}\hat{Z}^{\rho}_{b}Z_{*}^{b}h^{\beta\lambda}h^{\sigma\gamma}n^{\nu} \notag \\
    &-\epsilon_{\mu\nu\lambda\gamma}\hat{Z}^{\mu}_{c}\hat{Z}^{c\sigma}\hat{Z}^{\nu}_{b}Z_{*}^{b}h^{\alpha\rho}h^{\beta\gamma}n^{\lambda}-\epsilon_{\mu\nu\lambda\gamma}\hat{Z}^{\mu}_{c}\hat{Z}^{c\rho}\hat{Z}^{\nu}_{b}Z_{*}^{b}h^{\alpha\sigma}h^{\beta\gamma}n^{\lambda}-\epsilon_{\mu\nu\lambda\gamma}\hat{Z}^{\mu}_{c}\hat{Z}^{c\sigma}\hat{Z}^{\nu}_{b}Z_{*}^{b}h^{\alpha\gamma}h^{\beta\rho}n^{\lambda} \notag \\
    & -\epsilon_{\mu\nu\lambda\gamma}\hat{Z}^{\mu}_{c}\hat{Z}^{c\rho}\hat{Z}^{\nu}_{b}Z_{*}^{b}h^{\alpha\gamma}h^{\beta\sigma}n^{\lambda}-\epsilon_{\mu\nu\lambda\gamma}\hat{Z}^{\mu}_{c}\hat{Z}^{c\beta}\hat{Z}^{\nu}_{b}Z_{*}^{b}h^{\alpha\sigma}h^{\rho\gamma}n^{\lambda}-\epsilon_{\mu\nu\lambda\gamma}\hat{Z}^{\mu}_{c}\hat{Z}^{c\alpha}\hat{Z}^{\nu}_{b}Z_{*}^{b}h^{\beta\sigma}h^{\rho\gamma}n^{\lambda} \notag \\
    & -\epsilon_{\mu\nu\lambda\gamma}\hat{Z}^{\mu}_{c}\hat{Z}^{c\beta}\hat{Z}^{\nu}_{b}Z_{*}^{b}h^{\alpha\rho}h^{\sigma\gamma}n^{\lambda}-\epsilon_{\mu\nu\lambda\gamma}\hat{Z}^{\mu}_{c}\hat{Z}^{c\alpha}\hat{Z}^{\nu}_{b}Z_{*}^{b}h^{\beta\rho}h^{\sigma\gamma}n^{\lambda} \,.
\end{align}

Finally, after the change of basis in Sec. \ref{cb}, we get the structure given in \eq{Lkins}. For $\tilde{\mathcal{L}}_{4,2}^1 \Big|_{\rm{sl}}$, we get submatrices with zero in all the entries. Therefore, $\tilde{\mathcal{L}}_{4,2}^1 \Big|_{\rm{sl}}$ is trivially degenerate.



\section{Conclusions} \label{conclusions}
The GSU2P was built as a natural extension of the GP and, therefore, of the Horndeski theory to the case where the new gravitational degree of freedom is not only a vector field but one whose action enjoys a global SU(2) symmetry.  Its construction requires the implementation of the algebra constraint-enforcing relations to avoid the propagation of a fourth degree of freedom in the vector field.  It also requires its decoupling limit to be healthy, i.e., that it either produces second-order field equations for the longitudinal degrees of freedom or that the resulting action satisfies in turn a constraint algebra to avoid the propagation of unphysical scalar degrees of freedom.  Among the fourteen Lagrangian pieces for the GSU2P constructed in Ref. \cite{GallegoCadavid:2020dho}, the decoupling limits of eleven of them were very easy to analyze, leading to the conclusion that they either vanish or produce second-order field equations.  The other three were the subject of this paper because they produce higher-order field equations and, therefore, the analysis of their decoupling limits required the reconstruction of their kinetic matrices via a 3+1 ADM decomposition.  The results are satisfactory because the decoupling limits of these Lagrangian pieces turn out to satisfy the degeneracy condition, which is the primary constraint-enforcing relation. This, in particular, is very reassuring since the parity-conserving beyond SU(2) Proca terms play an important role in the successful constant-roll inflationary scenario studied in Ref. \cite{Garnica:2021fuu}.  Regarding the other members of the constraint algebra, it is not known whether they are satisfied. A definite answer to this question will only come with a dedicated and delicate Hamiltonian analysis, as was done for the DHOST in Ref. \cite{Langlois:2015skt}, which is not an easy task in curved spacetime.

\section*{Acknowledgments} 
 A. G. C. was funded by Agencia Nacional de Investigación y Desarrollo  ANID through the FONDECYT postdoctoral Grant No. 3210512. C. M. N. was supported by Vicerrectoría de Investigación y Extensión – Universidad Industrial de Santander
Postdoctoral Fellowship Programme No. 2021000126. Y. R. has received funding/support from the Patrimonio Autónomo - Fondo Nacional de Financiamiento para la Ciencia, la Tecnología y la Innovación Francisco José de Caldas (MINCIENCIAS - COLOMBIA) Grant No. 110685269447 RC-80740-465-2020, project 69553 as well as from the European Union’s Horizon 2020 research and innovation programme under the Marie Sklodowska-Curie grant agreement No 860881-HIDDeN. Some calculations were cross-checked with the Mathematica package xAct (www.xact.es).

\appendix
\section{Kinetic matrix entries of $\mathcal{L}_{4,2}^3 \Big|_{\rm{sl}}$}\label{aA}
The kinetic matrix entries in \eq{m1l1} are given by
\begin{equation}
     \tilde{\mathcal{C}}_{1} \equiv \sqrt{Z_{*}^{2}}\left(\hat{Z}^{2}+2(\hat{\tilde{Z}}_{1}^{\ 1})^{2}-3Z_{*}^{2}\right) \,,
\end{equation}
\begin{equation}
     \tilde{\mathcal{C}}_{2} \equiv 3\sqrt{2Z_{*}^{2}}\left(\hat{Z}^{2}+(\hat{\tilde{Z}}_{1}^{\ 1})^{2}-Z_{*}^{2}\right) \,,
\end{equation}
\begin{equation}
     \tilde{\mathcal{C}}_{3} \equiv 2(\hat{\tilde{Z}}_{1}^{\ 1})(\hat{\tilde{Z}}_{1}^{\ 2})\sqrt{Z_{*}^{2}} \,,
\end{equation}
\begin{equation}
     \tilde{\mathcal{C}}_{4} \equiv 3\sqrt{2Z_{*}^{2}}(\hat{\tilde{Z}}_{1}^{\ 1})(\hat{\tilde{Z}}_{1}^{\ 2}) \,,
\end{equation}
\begin{equation}
     \tilde{\mathcal{C}}_{5} \equiv 2(\hat{\tilde{Z}}_{1}^{\ 1})\sqrt{Z_{*}^{2}\left(\hat{Z}^{2}-(\hat{\tilde{Z}}_{1}^{\ 1})^{2}-(\hat{\tilde{Z}}_{1}^{\ 2})^{2}\right)} \,,
\end{equation}
\begin{equation}
     \tilde{\mathcal{C}}_{6} \equiv 3(\hat{\tilde{Z}}_{1}^{\ 1})\sqrt{2Z_{*}^{2}\left(\hat{Z}^{2}-(\hat{\tilde{Z}}_{1}^{\ 1})^{2}-(\hat{\tilde{Z}}_{1}^{\ 2})^{2}\right)} \,,
\end{equation}
\begin{equation}
     \tilde{\mathcal{F}}_{1} \equiv 6\hat{Z}^{4}-2\hat{Z}^{2}Z_{*}^{2}-4(\hat{\tilde{Z}}_{1}^{\ 1})^{2}Z_{*}^{2} \,,
\end{equation}
\begin{equation}
     \tilde{\mathcal{F}}_{2} \equiv \sqrt{2}Z_{*}^{2}\left(2\hat{Z}^{2}+7(\hat{\tilde{Z}}_{1}^{\ 1})^{2}-6Z_{*}^{2}\right) \,,
\end{equation}
\begin{equation}
     \tilde{\mathcal{F}}_{3} \equiv 3Z_{*}^{2}\left(\hat{Z}^{2}+(\hat{\tilde{Z}}_{1}^{\ 1})^{2}-2Z_{*}^{2}\right) \,,
\end{equation}
and those in \eq{m2l1} are given by
\begin{equation}
     \tilde{\mathcal{F}}_{4}\equiv 6\hat{Z}^{4}-3\hat{Z}^{2}Z_{*}^{2}-9(\hat{\tilde{Z}}_{1}^{\ 1})^{2}Z_{*}^{2}+6Z_{*}^{4} \,,
\end{equation}
\begin{equation}
     \tilde{\mathcal{F}}_{5} \equiv 3Z_{*}^{2}\left(-\hat{Z}^{2}-(\hat{\tilde{Z}}_{1}^{\ 1})^{2}+2Z_{*}^{2}\right) \,.
\end{equation}
\section{Kinetic matrix entries of $\mathcal{L}_{4,2}^4 \Big|_{\rm{sl}}$}\label{aB}
The kinetic matrix entries in \eq{m1l2} are given by
\begin{equation}
     \tilde{\mathcal{C}}_{7} \equiv 4\sqrt{Z_{*}^{2}}\left(\hat{Z}^{2}+2(\hat{\tilde{Z}}_{1}^{\ 1})^{2}-3Z_{*}^{2}\right) \,,
\end{equation}
\begin{equation}
     \tilde{\mathcal{C}}_{8} \equiv 4\sqrt{2Z_{*}^{2}}\left(2\hat{Z}^{2}+4(\hat{\tilde{Z}}_{1}^{\ 1})^{2}-3Z_{*}^{2}\right) \,,
\end{equation}
\begin{equation}
     \tilde{\mathcal{C}}_{9} \equiv 8\sqrt{Z_{*}^{2}}(\hat{\tilde{Z}}_{1}^{\ 1})(\hat{\tilde{Z}}_{1}^{\ 2}) \,,
\end{equation}
\begin{equation}
     \tilde{\mathcal{C}}_{10} \equiv 16\sqrt{2Z_{*}^{2}}(\hat{\tilde{Z}}_{1}^{\ 1})(\hat{\tilde{Z}}_{1}^{\ 2}) \,,
\end{equation}
\begin{equation}
     \tilde{\mathcal{C}}_{11} \equiv 8(\hat{\tilde{Z}}_{1}^{\ 1})\sqrt{Z_{*}^{2}\left(\hat{Z}^{2}-(\hat{\tilde{Z}}_{1}^{\ 1})^{2}-(\hat{\tilde{Z}}_{1}^{\ 2})^{2}\right)} \,,
\end{equation}
\begin{equation}
     \tilde{\mathcal{C}}_{12} \equiv 16(\hat{\tilde{Z}}_{1}^{\ 1})\sqrt{2Z_{*}^{2}\left(\hat{Z}^{2}-(\hat{\tilde{Z}}_{1}^{\ 1})^{2}-(\hat{\tilde{Z}}_{1}^{\ 2})^{2}\right)} \,,
\end{equation}
\begin{equation}
     \tilde{\mathcal{F}}_{6} \equiv 8\left(3\hat{Z}^{4}-\hat{Z}^{2}Z_{*}^{2}-(\hat{\tilde{Z}}_{1}^{\ 1})^{2}Z_{*}^{2}\right) \,,
\end{equation}
\begin{equation}
     \tilde{\mathcal{F}}_{7} \equiv 12\sqrt{2}Z_{*}^{2}\left(\hat{Z}^{2}+2(\hat{\tilde{Z}}_{1}^{\ 1})^{2}-2Z_{*}^{2}\right) \,,
\end{equation}
\begin{equation}
     \tilde{\mathcal{F}}_{8} \equiv 8Z_{*}^{2}\left(\hat{Z}^{2}+2(\hat{\tilde{Z}}_{1}^{\ 1})^{2}-3Z_{*}^{2}\right) \,,
\end{equation}
and those in \eq{m2l2} are given by
\begin{equation}
     \tilde{\mathcal{F}}_{9} \equiv 8\left(3\hat{Z}^{4}-2\hat{Z}^{2}Z_{*}^{2}+Z_{*}^{2}\left(-4(\hat{\tilde{Z}}_{1}^{\ 1})^{2}+3Z_{*}^{2}\right)\right) \,,
\end{equation}
\begin{equation}
     \tilde{\mathcal{F}}_{10} \equiv -8Z_{*}^{2}\left(\hat{Z}^{2}+2(\hat{\tilde{Z}}_{1}^{\ 1})^{2}-3Z_{*}^{2}\right) \,.
\end{equation}

\bibliography{Bibli.bib} 

\end{document}